# IoT - Based Traffic Management System for Ambulances

Mohammad Moazum Wani
Department of Information Technology
Central University of Kashmir
Ganderbal J&K , India
moazumwani21@gmail.com

Samiya Khan
Department of Computer Science
Jamia Millia Islamia
New Delhi, India
samiyashaukat@yahoo.com

Mansaf Alam
Department of Computer Science
Jamia Millia Islamia
New Delhi, India
malam2@jmi.ac.in

*Abstract*— **Lack of efficient traffic control can lead to the loss of thousands of lives due to ambulance not being able to reach the hospital in time. Also, with the current annual growth of vehicles being around 11% while the annual road extension remaining around 4% in developing countries such as India, the problem is further worsening. So, to deal with this problem the paper presents a novel, easy to implement alternative for traffic management during emergency situations requiring only three main devices: Arduino UNO, GPS neo 6M and SIM 900A.**

## I. INTRODUCTION

One of the most profound aftermaths of evolving technologies in this modern era is rapidly increasing vehicular counts, which has become grave in the wake of staggering rise in world population. As a result, traffic congestion has become a serious problem in most countries around the world. Moreover, the rate of road expansion is just one-third the vehicular growth rate in countries like India, which further worsens the problem [1]. Due to increased traffic, there has been an increase in traffic jams, which has caused a problem for ambulances in moving patients to the appropriate destinations on time in case of emergency. Statistically, it has been observed that more than 20% patients requiring urgent medical attention die on their way to the hospital because of delays [2].

The International Road Federation, Geneva Programme Centre has reported 3,000 deaths everyday as a result of road accidents. This approximates to a yearly figure of 1.3 million and a total count of 2.4 million road accident deaths [3]. One of the identified reasons for deaths is delayed emergencies services and the ambulance unable to reach the accident location and then, to the hospital, in time. A recent case that validates the gravity of the situation occurred in Rohtak, India where a child died soon after reaching PGIMS as the ambulance was held up in traffic for close to half an hour [4].

Another similar incident happened in Mysore, India where a 2-year old girl was rushed to Manipal hospital after her blood pressure dropped and she was unable to breathe. The ambulance took 2.5 hours to cover the distance of 15 km. She was declared 'brought dead' by the hospital [5]. This problem is known to exist in similar and greater dimensions, in other countries as well. In Manila, heavy traffic kept the ambulance trapped for an extra 10 minutes as the result of which the patient died. The hospital was just 5.7 km away from the site of pickup. Recently, a woman shared a video over the Internet expressing her astonishment on how cars on the way did not give way to the ambulance in which she was carrying her mother for treatment to the nearby hospital [6].

However, the problem of an ambulance not reaching its desired location is not being solved yet. So, in order to deal with this problem, we need a system, in which the ambulance does not need to stop anywhere on the way and can directly go to its destination. This research paper proposes an IoT based traffic management system for an ambulance. The proposed work is hardware-based and uses Arduino UNO [7], GSM SIM 900A and GPS neo 6M.

The rest of the paper is organized in the following manner. Section II identifies the existing solutions, their limitations and scope for future research in this field. Section III describes the methodology used for design and development of the proposed solution. The experimental results are summarized in Section IV and the analysis of the results is provided in Section V. Lastly, Section VI summarizes the findings of this work, elaborating on the scope of future research.

## II. LITERATURE REVIEW

Ahir et al. [8] proposed an android application and hardware module for traffic signal implementation. The android application has four buttons for four directions. It depends on the route which ambulance driver will select, the ambulance driver will select the appropriate direction and send activate command for the particular signal. In the android application, the patient's information is also stored which consists of name, age, and blood group, besides others. The hardware module for traffic signal has used Arduino for traffic signals. It consists of a Wi-Fi module, with the help of which it captures information from the server and connects the android application directly with the traffic signal.

The location is retrieved in the form of longitude and latitude. The direction of ambulance movement depends on the degree of the ambulance. For detecting the direction and current location of the ambulance, a compass is used.

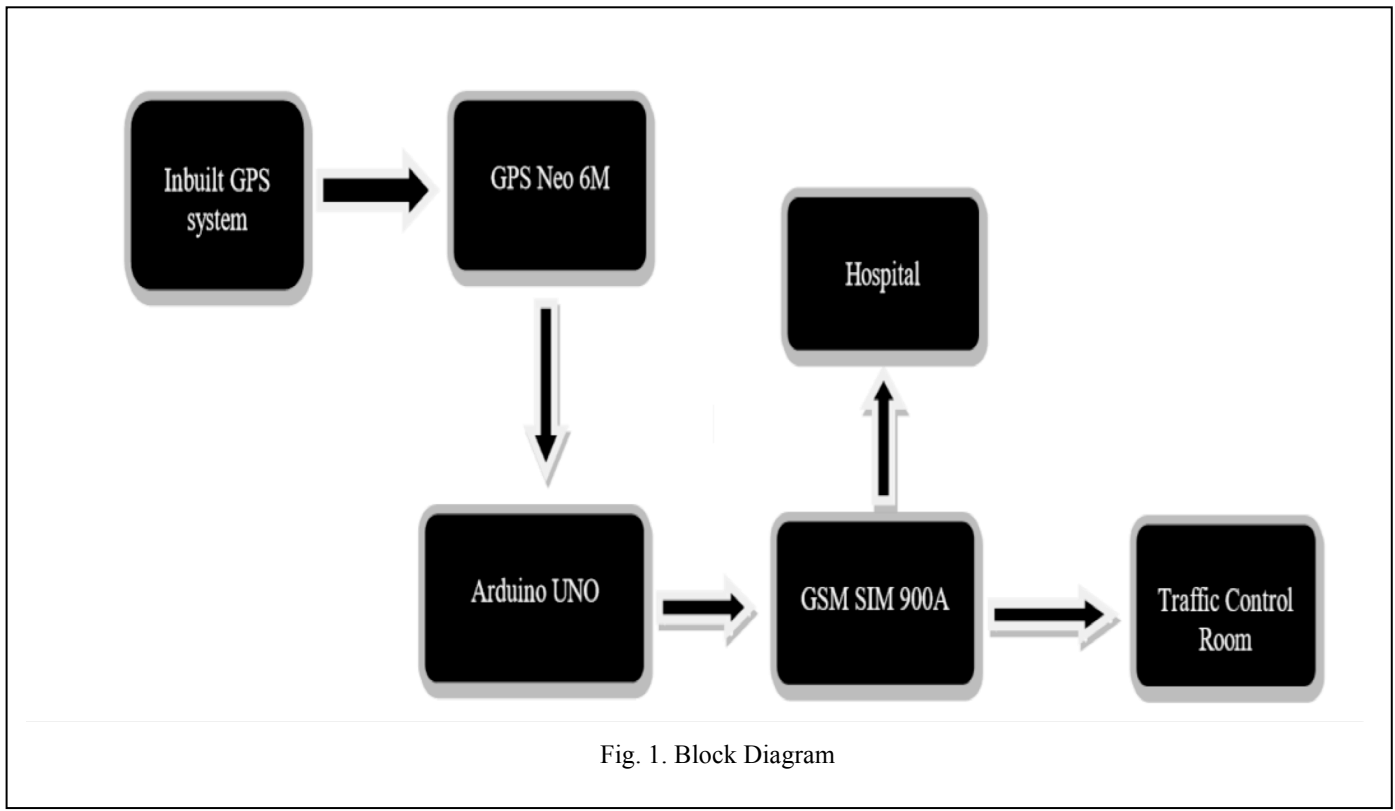

Fig. 1. Block Diagram

The system used in this paper [8] in manual, neither does it does not give the shortest path to the hospital nor does the signal changes automatically.

Parida et al. [9] uses an RF transmitter-receiver module and a Zigbee transmitter-receiver along with Atmega 328 IS's. This system enables emergency vehicles to override current traffic light sequence and reach its destination uninterrupted. The whole system depends on communication between emergency vehicles and traffic signals. The RF transmitter-receiver module and a Zigbee transmitter-receiver are both connected to the microcontroller.

There could be a case when there is a traffic jam caused by minor accident or road construction, which will not be known to the emergency vehicle driver. So in that case, this system is of no use to the emergency vehicle driver, as the emergency vehicle will still be stuck in a traffic jam even after having control of traffic light. The hardware used in this system [9] is short-ranged and have low power.

Kandhari and Antonov [10] proposed a system, which segments the overall system into three phases, which include counting and detection of vehicles, detection of ambulance and counting and data-based decision makings based on data. The system made use of a detection mechanism, which based its decision on images. Therefore, this system does not use electronic sensors. The route is cleared for emergency vehicles stuck in traffic using Bluetooth technology. Bluetooth module and a Bluetooth-enabled phone with Bluetooth state active are required for the detection phase.

For this system to work, the ambulance should be near the signal for sending the command for guiding the traffic signal to the Bluetooth module. The range of Bluetooth is very low so there can be a case that the driver sends the command using his phone and it was not received by the Bluetooth module. This system [10] suffers from security-related issues. In order to make it secure the code needs to be changed every 24 hours.

The system proposed by Ramani and Jeyakumar [11] makes use of a central server for controlling the traffic controllers. Arduino UNO is used for the implementation of the traffic signal controller. In this system [11], a web application is provided to the ambulance driver. The request for turning the signal light to green is sent to the traffic controller using the web application. The system is divided into three parts namely, web application, cloud server and traffic signal.

The ambulance driver uses the web application for choosing the route and navigating the ambulance. Communication between the ambulance and traffic signal is established using the cloud server. Arduino UNO is interfaced with Wi-Fi module and the Wi-Fi module is used as a traffic signal in this system.

This system always requires sending of a signal to turn the traffic signal green, as it is not automatic. There could be a case in which the driver forgets to send the signal because of the time constrains. In that case, the system will not work. For this system to work properly, Internet should be available to the drivers all the time, which is not possible in all the areas.

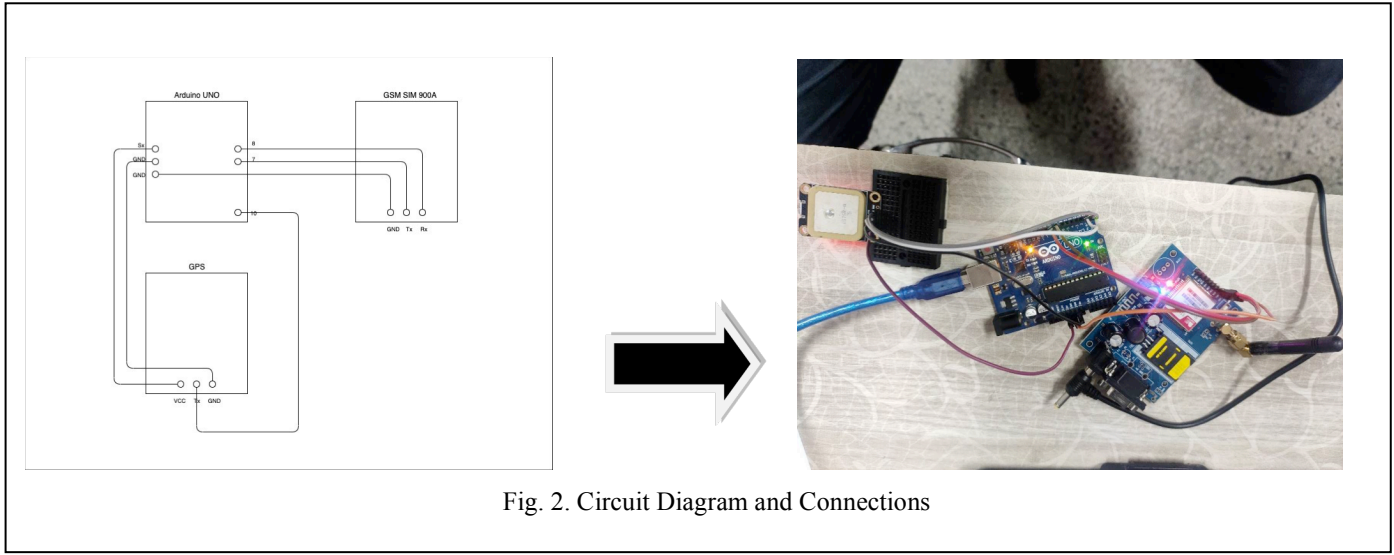

Fig. 2. Circuit Diagram and Connections

Bondade et al. [12] uses three modules namely, admin, driver/ambulance and controlling signal. This system [12] is actually designed to monitor the density of traffic on the road. According to this system, if there is an ambulance on the road, then the image processing system will detect the ambulance and will inform the system to clear the traffic in critical condition by turning the traffic light green, allowing traffic, for the path on which the emergency vehicle is travelling and all other signals will be turned to red, disallowing traffic.

The admin module is used to provide a route to the driver of the ambulance and it also tracks the source and destination of the ambulance. The driver /ambulance module is used for sending the request to the admin for permission to reach the location. The controlling signal module is used to manipulate traffic signals.

Problem is expected to occur if there are two ambulances coming from a different route. In that case, the system will allow only one ambulance and the other will be stuck in the red light. The receiver randomly selects the ambulance without consideration of any parameters. The ambulance driver always has to send a request to the admin to give them permission to reach the location, and then the driver also has to wait for the admin to send the route for the ambulance driver.

Existing solutions make use of Bluetooth or Internet-based platforms which pose major limitations in terms of connectivity and range-related issues. Moreover, the use of RF Transmitters has a range and power limitations which can be handled by expensive hardware replacements. The connected solution has an inherent security problem associated with them. Besides these, some specific issues like requirement of admin permission for route management and handling of situations involving multiple ambulances, exist in surveyed solutions. The proposed system mitigates existing limitations, in entirety.

III. METHODOLOGY

The scope of this project is in cities with huge traffic, which will make it difficult for an ambulance to arrive on time. The block diagram of the proposed system is provided in Fig. 1 This project consists of four main hardware components:

A. Inbuilt GPS system

B. GPS Module NEO 6M

C. Arduino UNO

D. SIM 900A GSM Modem

Descriptions of the hardware components are provided below. In addition to the above-mentioned, the system also includes subcomponent named Traffic Control Room. The traffic control room will help the ambulance to reach its destination on time by clearing the route for the ambulance and by changing the traffic signals whenever and wherever needed. Algorithm of the code for the proposed system is provided in Algorithm 1.

*A. Inbuilt GPS System*

Google Maps [13] is to be installed in the inbuild GPS system present in the ambulance. In the google maps, the location of all the hospitals and nursing homes will be stored in the Google Maps interface. The GPS will select the shortest possible path to reach the nearest hospital. The inbuilt GPS of the ambulance will be connected to the GPS Neo 6M.

*B. GPS Module NEO*

GPS module NEO 6M sends live location of the ambulance to the traffic control room and the hospital. So, the traffic control room can accordingly clear a route for the ambulance.

*C. Arduino UNO*

Arduino UNO [7] is used to store the code for sending the live location of the ambulance. It receives the location from the GPS Neo 6M and sends it to the traffic control room and hospital using SIM 900A.

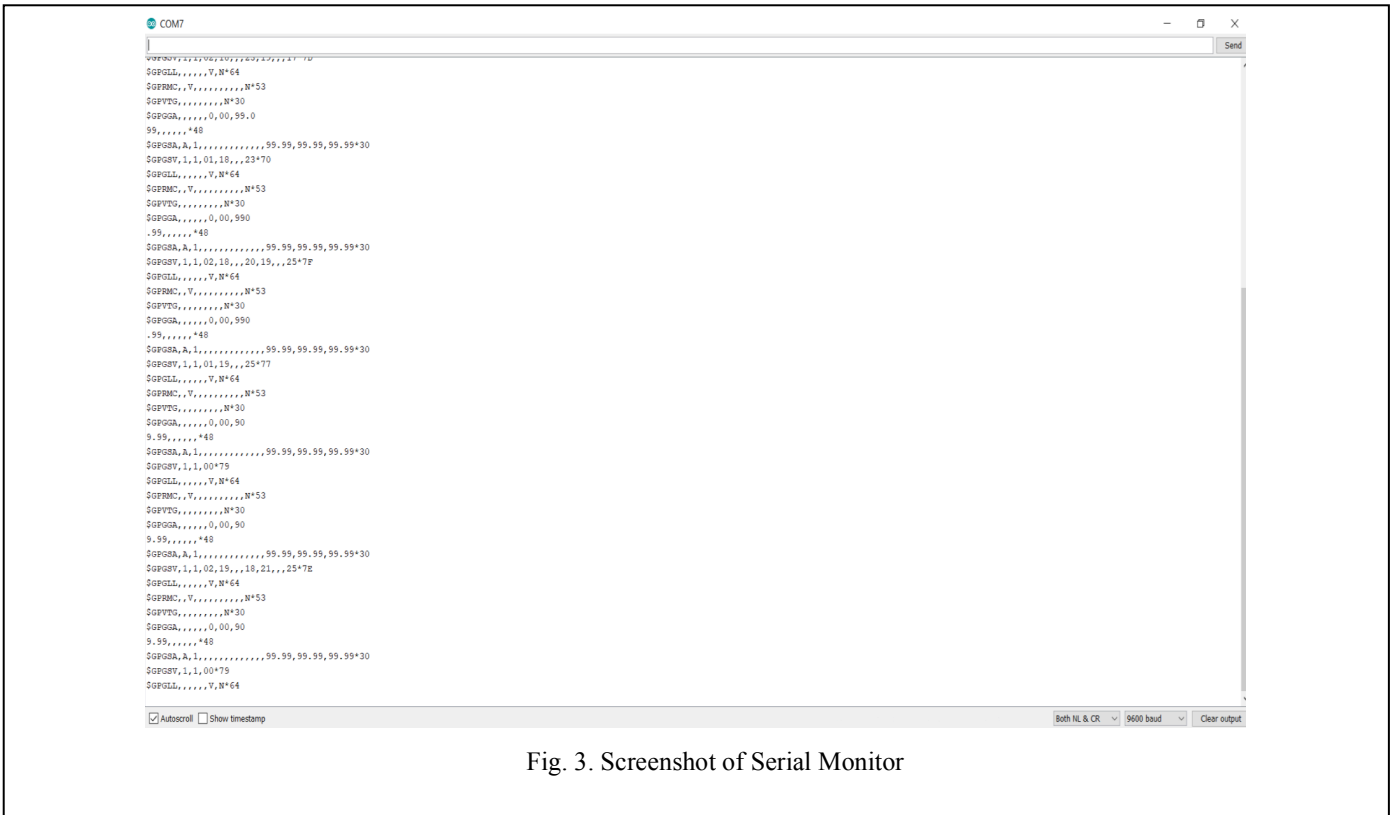

Fig. 3. Screenshot of Serial Monitor

*D. GSM SIM 900A*

SIM 900A is a GSM enabled SIM, which is used to send live location of the ambulance using text message to the traffic control room and hospital.

IV. EXPERIMENTAL VALIDATION

The proposed solution has been implemented and tested using a prototype the connects of the Arduino-based solution are illustrated in Fig. 2. Once the system is integrated with an ambulance the driver can choose the destination hospital. The link of the live location will be automatically sent to the traffic control room and the hospital. The Google Maps will provide the shortest path from the source to the destination hospital and the traffic control room will clear the traffic on the route.

Fig. 2 is the circuit of the system. There are three connections between GSM SIM 900A and Arduino UNO. Rx of GSM SIM 900A with digital pin 8(Tx) of Arduino UNO: Tx of GSM SIM 900A with digital pin 7(Rx) of Arduino UNO and GND of GSM SIM 900A with GND of Arduino UNO. There are also three connections between GPS Neo 6M and Arduino UNO: GND to GPS with GND of Arduino, VCC of GPS with 5V of Arduino UNO and Tx of GPS with digital pin 0 (Rx) of Arduino UNO. Arduino UNO is connected to the computer so that we can upload the code in the system. The power supply of 12V, 1A is applied to the GSM board for power supply.

The serial monitor is illustrated in Fig. 3 and shows the continuous location of the GPS neo 6M. The serial monitor helps the system to check whether GPS is working or not. The message sent by GSM SIM 900A includes the location of the ambulance at the starting point, the location, which can be continuously monitored by the traffic control room and hospital. A click on the Google Maps link opens the live location of the ambulance. Fig. 4 provides a graphical illustration of this functionality.

V. DISCUSSION

This research work proposes a semi-automatic system for managing traffic control for healthcare-related emergencies.

---

**ALGORITHM 1**

1. Initialize variables: *newData = false*
2. For GPS Parsing Time Lapsed < 1 second
3. *IF serial connection is available*
4. Read data from serial connection
5. *ENDIF*
6. *IF data is read*
7. *newData = true*
8. *ENDIF*
9. *IF newData = true*
10. Identify longitude and latitude of the current location of the ambulance
11. Generate Google Maps link for the location
12. Send message
13. *ENDIF*

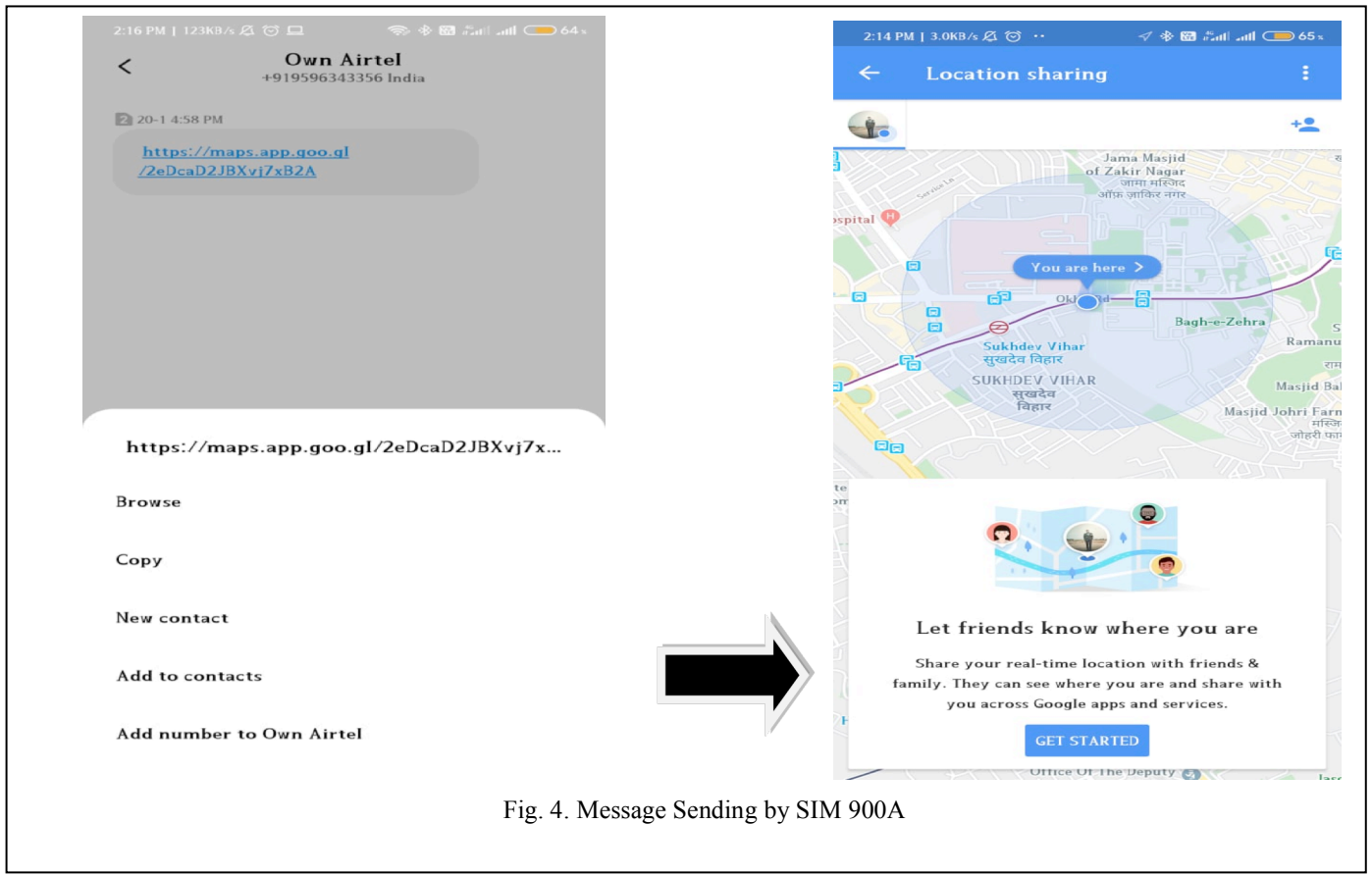

Fig. 4. Message Sending by SIM 900A

The GPS will show the shortest path to the concerned destination and the live location will be sent to the hospital and traffic control room. The traffic control room will clear the traffic on the route and traffic signal management will be done accordingly.

The proposed system can be easily integrated into the ambulance as it only needs 12V,1A power for GSM SIM 900A and 10V for Arduino UNO, which can be easily provided from the fuse board present that is inside the ambulance. The proposed system in existing literature needed that driver to have an Internet connection. In the proposed system, the driver just needs to click on the GPS screen once. There is a need to send the location of the ambulance as a message continuously. This must be done once because the location that is sent from device acts as a live location. The proposed approach can give way to one or more ambulance at the same time.

## VI. Conclusion

This research paper proposes an Arduino based traffic control system for healthcare-related emergencies. All though, the system is expected to work well on its base functionality, it suffers from hardware-related limitations. The connections of the system should be made carefully. If there is a mistake in joining the connections the system will not work properly. Future scope of this research includes the integration of the proposed system to sensor-based patient data collection modules. The data will be sent to the Cloud using Arduino-based Wi-Fi module. The destination hospital can access the real-time patient data using open Wi-Fi system. The proposed system can be improved in this direction for future use.